\documentclass[twocolumn,aps,prd,showpacs]{revtex4}

\usepackage{graphicx}
\usepackage[usenames]{color}
\usepackage{psfrag}
\usepackage[ps2pdf,colorlinks,bookmarks]{hyperref}
\usepackage{bm}

\definecolor{linkblue}{rgb}{0,0,0.8}
\definecolor{linkgreen}{rgb}{0,0.5,0}
\hypersetup{pdfpagemode=None, pdfstartview=FitH, linkcolor=linkblue, %
            citecolor=linkgreen, urlcolor=linkblue}

\newcommand{\ud}[2]{^{#1}_{\phantom{#1}#2}}

\def\beq{\begin{equation}}
\def\eeq{\end{equation}}
\def\bea{\setlength\arraycolsep{1.4pt}\begin{eqnarray}}
\def\eea{\end{eqnarray}}
\def\bit{\begin{itemize}}
\def\eit{\end{itemize}}
\def\nn{\nonumber}

\def\eq{Eq.~}
\def\eqs{Eqs.~}

\def\ld{\left}
\def\rd{\right}

\def\fr{\frac}
\def\oo{\frac{1}}
\def\half{\frac{1}{2}}

\def\del{\delta}

\def\lam{\lambda}
\def\Lam{\Lambda}

\def\Om{\Omega}

\def\p{{\cal P}}
\def\R{{\cal R}}

\def\bra{\langle}
\def\ket{\rangle}

\def\alm{a_{\ell m}}
\def\Ylm{Y_{\ell m}}

\def\rls{r_{\rm LS}}

\def\LCDM{$\Lambda$CDM}
\def\Planck{\textit{Planck}}

\newcommand{\sYlm}[3]{{}_{#1}Y_{#2#3}}
\def\kv{\mathbf{k}}
\def\kh{\hat{\mathbf{k}}}

\begin{document}

\title{Scalar-tensor correlations and large-scale power suppression}

\author{James P. Zibin} 
\email{zibin@phas.ubc.ca}
\affiliation{Department of Physics and Astronomy, %
University of British Columbia, %
Vancouver, British Columbia, V6T 1Z1  Canada}

\date{\today}

\begin{abstract}

   Recent measurements from the BICEP2 cosmic microwave background 
polarization experiment indicate the presence of primordial gravitational 
waves with surprisingly large amplitude.  If these results are confirmed, 
they point to a discrepancy with temperature anisotropy power spectrum 
measurements and suggest that extensions to the standard cosmological model 
may be required to resolve the discrepancy.  One 
intriguing extension is an anticorrelation between tensors and 
scalars to naturally suppress the temperature power.  Here I examine 
this possibility and show that such a suppression is not possible in 
the presence of a general form of anticorrelation.

\end{abstract}


\maketitle

\section{Introduction}
\label{introsec}

   The standard $\Lam$ cold dark matter (\LCDM) model has been remarkably 
successful at describing the large scale geometry, content, and thermal 
history of the Universe.  Nevertheless, whispers of discrepancies from 
the standard model at the largest observable scales have been heard.  In 
particular, there is a deficit of temperature anisotropy power at the 
largest angular scales in the cosmic microwave background (CMB) (with 
respect to the best-fit six parameter model)~\cite{planckresults}.  
In addition, a roughly dipolar power asymmetry 
is present on multipole scales $\ell \lesssim 100$ (see, e.g., 
\cite{planck13isostat}).  These features are not of very high significance 
and are sensitive to {\em a posteriori} choices, so they may simply 
turn out to be the result of large Gaussian random field fluctuations.  
Nevertheless, they have attracted considerable attention as they may be 
hinting at extensions to \LCDM.

   The whispers of discrepancies have turned into shouts recently with the 
announcement of results from the BICEP2 CMB polarization 
experiment~\cite{bicep2}.  The BICEP team has performed a measurement of 
the $B$-mode polarization power spectrum and concluded that their results 
indicate the presence of a remarkably large-amplitude primordial gravitational 
wave power spectrum, with tensor-to-scalar ratio $r = 0.16^{+0.06}_{-0.05}$.  
This result must pass a number of tests before this conclusion can be widely 
accepted.  First, it must be determined whether polarized galactic 
foreground emission or other systematic effects might account for the 
signal (see, e.g., \cite{lms14}).  Next, the possibility that the signal, 
if extragalactic, is due to something other than primordial gravitational 
waves, such as defects, magnetic fields, or birefringence, must be 
considered~\cite{mp14,lizarragaetal14,bdm14,lln14}.  Nevertheless, the BICEP 
announcement potentially represents one of the most important discoveries in 
the history of cosmology and should be taken very seriously.

   One particularly surprising aspect of the BICEP result is the apparent 
discrepancy with CMB temperature power spectrum measurements.  The \Planck\ 
satellite recently placed a 95\% upper limit of 
$r < 0.11$~\cite{planckparams}.  Such temperature power limits are based 
on the characteristic shape of the tensor temperature power spectrum, 
namely, that of a large-scale plateau which tapers off on scales smaller 
than $\ell \sim 100$.  The absence of any visible large-scale power excess 
limits the possible tensor contribution to within cosmic variance.  
The fact that we actually observe a large-scale power {\em deficit} only 
exacerbates this discrepancy, raising it to approximately the $3\sigma$ 
level~\cite{smithetal14}.

   Many approaches to resolving this discrepancy, based on the assumption 
that the BICEP measurement is correct, are possible.  Generically, they 
require the introduction of extra cosmological parameters which have the 
effect of suppressing temperature power on large scales, to compensate for 
a large tensor contribution.  Perhaps the simplest possibility is the 
introduction of a negative running of the primordial scalar power spectrum 
tilt, as pointed out by the BICEP team themselves~\cite{bicep2}.  However, 
the required running is much larger than that expected in the simplest 
inflationary models (see, e.g., \cite{planckinfl}).  Other possibilities 
include the addition of an anticorrelated isocurvature 
component~\cite{ksty14} or of additional neutrino species (see, e.g., 
\cite{archidiaconoetal14}).  Of course, an {\em ad hoc} procedure of 
suppressing the primordial scalar power on the largest scales is certainly a 
possibility.  While inflationary models with such features have been 
discussed (see, e.g, \cite{hsss14,hsss14b,adss14}), they require an amplitude 
and cutoff scale tuned remarkably to coincide with and compensate for the 
tensor contribution.

   One intriguing possibility is that of an anticorrelation between 
tensors and scalars, which might offer the possibility of naturally 
suppressing temperature power without the need to introduce a scale or 
an amplitude by hand~\cite{cps14}.  Of course, such correlations do not 
occur in the simplest models of inflation, and so would necessitate the 
introduction of complications to the basic models (see, e.g., \cite{ghp10}).  
Nevertheless, it is worth investigating 
the viability of this approach.  In this brief report I attempt to address 
the question of how well temperature power can be reduced with a general 
form of tensor-scalar correlation.  I calculate the total temperature 
anisotropy power due to tensors and scalars on large angular scales in 
the presence of such a correlation.  I find that a reduction to the 
total measured temperature power is not possible, in agreement with a 
special case analyzed very recently in~\cite{cefw14}.

\section{Scalar-tensor correlations}


   In this section, I consider the most general form that a correlation can 
take between tensor modes, represented by the transverse-traceless (TT) 
spatial metric perturbation $h_{ij}$, and scalar modes, represented here by 
the comoving curvature perturbation $\R$.  Both scalar and tensor modes are 
described in this section by their primordial values, i.e.\ their values on 
super-Hubble 
scales after inflation, and hence can be treated as time independent in 
ordinary adiabatic models.  Time dependence will be straightforwardly 
implemented later in the analysis.

   Scalar modes are taken to satisfy the two-point correlation function
\beq
\bra\R^*(\kv)\R(\kv')\ket = \fr{2\pi^2}{k^3}\p_\R(k)\del^3(\kv - \kv'),
\eeq
with dimensionless power spectrum $\p_\R(k)$ and comoving wave vector $\kv$.  
For the tensors, if we use the standard helicity states
\beq
e_{ij}^{\pm 2}(\hat{\mathbf{k}}) = \oo{\sqrt{2}}\ld(e_{ij}^+(\hat{\mathbf{k}})
                                \pm ie_{ij}^\times(\hat{\mathbf{k}})\rd),
\eeq
we can expand the modes according to
\beq
h_{ij}(\mathbf{k}) = h_\lam(\mathbf{k})e_{ij}^\lam(\hat{\mathbf{k}}),
\eeq
where repeated helicity indices $\lam = \pm2$ are summed over.  Then we 
can write the two-point correlator as
\beq
\bra h_\lam^*(\mathbf{k})h_{\lam'}(\mathbf{k}')\ket
   = \fr{\pi^2}{2k^3}\p_h(k)\del^3(\mathbf{k} - \mathbf{k}')\del_{\lam\lam'},
\label{hlamPS}
\eeq
with dimensionless tensor power spectrum $\p_h(k)$.  The tensor-to-scalar 
ratio at pivot scale $k_*$ is then defined by $r \equiv \p_h(k_*)/\p_\R(k_*)$.

   The TT character of the tensor perturbations implies that a tensor-scalar 
correlation must take the form
\beq
\bra h_{ij}^*(\mathbf{k})\R(\mathbf{k}')\ket
   = \fr{2\pi^2}{k^3}\p_{h\R}(k)C_{ij}(\hat{\mathbf{k}})
     \del^3(\mathbf{k} - \mathbf{k}'),
\label{hijcorn}
\eeq
where the (otherwise arbitrary) spatial tensor $C_{ij}(\hat{\mathbf{k}})$ 
satisfies
\beq
k^iC_{ij} = C\ud{i}{i} = 0, \quad C_{ij} = C_{ji},
\label{Ctt}
\eeq
and $\p_{h\R}(k)$ is the dimensionless correlated power spectrum.  Crucially, 
no tensor satisfying \eq(\ref{Ctt}) can be constructed from the metric or 
wave vector $\mathbf{k}$, so $C_{ij}$ must correspond to a new tensor field 
or to a breaking of Lorentz invariance.  In terms of the helicity states 
this correlation becomes
\beq
\bra h_\lam^*(\mathbf{k})\R(\mathbf{k}')\ket
   = \fr{\pi^2}{k^3}\p_{h\R}(k)D_{ij}e^{ij}_\lam(\hat{\mathbf{k}})
     \del^3(\mathbf{k} - \mathbf{k}'),
\eeq
for arbitrary tensor $D_{ij}$.  The tensor $C_{ij}$ is related to $D_{ij}$ via 
a projection into the TT subspace, i.e.
\beq
C^{kl}(\hat{\mathbf{k}})
   = \half D^{ij}e_{ij}^\lam(\hat{\mathbf{k}})e^{kl}_\lam(\hat{\mathbf{k}}).
\eeq
The tensor $D_{ij}$ describes the correlation more fundamentally than 
$C_{ij}(\hat{\mathbf{k}})$, since the former is independent of the mode 
direction.

   Relation (\ref{hijcorn}) implies that the general tensor-scalar 
correlation for each wave vector depends on two independent parameters, the 
unique components of $C_{ij}$.  But the physical correlation tensor $D_{ij}$ 
depends effectively on five parameters, since an isotropic (trace) part 
of $D_{ij}$ does not contribute to correlations.  Two of those five parameters 
can be taken to determine the orientation of the first principle axis 
of $D_{ij}$, and one more parameter describes the orientation of the second 
principle axis in the plane orthogonal to the first.  Therefore, up to 
overall rotations, there are only two parameters in $D_{ij}$, describing, 
e.g., the lengths of the second and third principle axes relative to the 
first.  Choosing the coordinate and principle axes to coincide, the most 
general correlation takes the form
\beq
D^{ij} = \alpha\hat{x}^i\hat{x}^j + \beta\hat{y}^i\hat{y}^j
       + \hat{z}^i\hat{z}^j,
\eeq
for constants $\alpha$ and $\beta$.

   We can easily evaluate the correlations explicitly in the case $\alpha = 
\beta = 0$.  Then we have
\beq
D_{ij}e^{ij}_\lam(\hat{\mathbf{k}})
   = \hat{z}_i\hat{z}_j e^{ij}_\lam(\hat{\mathbf{k}})
   = \hat{z}^i\hat{z}^j S_{ik}(\hat{\mathbf{k}})S_{jl}(\hat{\mathbf{k}})
     e^{kl}_\lam(\hat{\mathbf{z}}),
\eeq
where $S_{ij}(\hat{\mathbf{k}})$ represents a standard rotation from the 
$\hat{\mathbf{z}}$ to $\hat{\mathbf{k}}$ directions.  Writing 
$\hat{\mathbf{k}} = (\theta_k,\phi_k)$ and using the explicit form for the 
rotation matrices, we find
\beq
D_{ij} e^{ij}_\lam(\hat{\mathbf{k}})
   = \oo{\sqrt{2}}\sin^2(\theta_k)
   = 4\sqrt{\fr{\pi}{15}}\sYlm{\lam}{2}{0}(\hat{\mathbf{k}}),
\label{zze}
\eeq
where the $\sYlm{\lam}{\ell}{m}$ are the spin-$\lam$ spherical harmonics.  
As expected, the correlation behaves like a quadrupolar spin-$\lam$ 
quantity.  More general correlations with nonzero $\alpha$ and $\beta$ 
will entail mixtures of $\sYlm{\lam}{2}{m}$ with all $|m| \le 2$, 
although we will not need their explicit forms.

   Before concluding this section, note that positivity of total power 
puts a restriction on the magnitude of a tensor-scalar anticorrelation.  
In particular, we must have
\bea
\!\!\!\!\bra|\gamma h_\lam(\kv) + \R(\kv)|^2\ket 
   &\propto& |\gamma|^2\p_h(k) + 4\p_\R(k)\nn\\
       && + 4\mathrm{Re}\ld(\gamma^*\p_{h\R}(k)D_{ij}
            e^{ij}_\lam(\hat{\mathbf{k}})\rd)\\
   &\ge& 0
\eea
for any $\gamma$.  This effectively puts a constraint on the magnitude 
of the correlated power, $\p_{h\R}(k)$, for the case of anticorrelations.  
This correlated power was apparently treated as a free function 
in~\cite{cps14}.

\section{CMB temperature anisotropies}

   As mentioned in the Introduction, in order to determine whether 
tensor-scalar anticorrelation can reduce large-angle temperature anisotropy 
power we must use the general correlation described above to calculate the 
CMB anisotropies.  In this section, I will explicitly calculate the scalar 
Sachs-Wolfe (SW) effect anisotropies, which is a reasonable approximation to 
the total scalar power on large scales, where the tensor contribution is 
significant.  Similarly, I will calculate the anisotropies due to the line 
of sight tensor integrated SW effect.  This is a very good approximation, 
and only ignores effects due to noninstantaneous recombination and 
neutrino damping.  Importantly, Ref.~\cite{cps14} approximated the tensor 
contribution as localized to the last scattering surface.  In fact, the 
contribution is relatively broadly distributed along the line of sight, 
which means that the correlation is expected to be weaker than predicted 
in \cite{cps14}.

   The scalar SW temperature anisotropy is simply
\beq
\fr{\del T^S(\hat{\mathbf{n}})}{T} = -\oo{5}\R(\rls\hat{\mathbf{n}}),
\eeq
where $\rls$ is the comoving radius to last scattering.  Decomposing as usual 
into spherical harmonics, $\del T(\hat{\mathbf{n}})/T = 
\sum_{\ell m}\alm\Ylm(\hat{\mathbf{n}})$, we find for the scalar multipole 
coefficients
\beq
\alm^S = -\oo{5}\sqrt{\fr{2}{\pi}}i^\ell
         \int dk k^2j_\ell(k\rls)\int d\Om_k \R(\kv)\Ylm^*(\kh),
\label{almS}
\eeq
for spherical Bessel function $j_\ell$.  The scalar power is easily 
calculated to be
\beq
C_\ell^S = \bra\alm^{S*}\alm^S\ket
         = \fr{4\pi}{25}\int\fr{dk}{k}\p_\R(k)j_\ell^2(k\rls),
\label{ClS}
\eeq
which describes the familiar nearly flat SW plateau.

   The line of sight temperature anisotropy due to tensors is
\beq
\fr{\del T^T(\hat{\mathbf{n}})}{T}
   = -\half\int_E^R\dot{h}_{ij}\hat{n}^i\hat{n}^j dt.
\label{delTtens}
\eeq
Here $E$ and $R$ represent the emission point on the last scattering surface 
and the reception point at the origin today, respectively.  Ignoring the 
effects of neutrino and photon anisotropic stress, the tensor 
fluctuations evolve according to the linearized Einstein equation
\beq
\ddot{h}_{ij} + 3H\dot{h}_{ij} - \fr{\nabla^2}{a^2}h_{ij} = 0,
\eeq
with Hubble rate $H \equiv \dot{a}/a$.  These perturbations evolve very 
slowly at early times when the modes are super-Hubble, but start to decay 
as the modes cross the Hubble radius.  Thus the anisotropy 
\eq(\ref{delTtens}) on multipole scale $\ell$ is sourced mainly by modes 
with $k \sim \ell/r \sim a(r)H(r)$; i.e., it is sourced along a substantial 
interval of the line of sight.  Since we only observe to a maximum distance 
$\rls$, we can see that the tensor temperature anisotropies are only 
sourced on scales larger than $\ell \sim \rls a(\rls)H(\rls) \sim 100$.

   Expanding into helicity modes, we can factor the late-time evolution 
out from the primordial amplitudes by defining
\beq
h_\lam(\kv,t) = h(k,t)h_\lam(\kv),
\eeq
with the number of arguments differentiating the functions $h_\lam(\kv,t)$ 
and $h_\lam(\kv)$ and with the time-dependent factor satisfying
\beq
\ddot{h}(k,t) + 3H\dot{h}(k,t) + \fr{k^2}{a^2}h(k,t) = 0.
\eeq

   Expanding again into spherical harmonics, we find after some computation
\bea
\alm^T &=& \fr{i^\ell}{2\sqrt{\pi}}\ld[\fr{(\ell + 2)!}{(\ell - 2)!}\rd]^{1/2}
       \int dk k^2 f_\ell(k)\nn\\
     &\times& \int d\Om_k h_\lam(\kv)\sYlm{\lam}{\ell}{m}^*(\kh),
\label{almT}
\eea
where
\beq
f_\ell(k) \equiv \int_E^R \dot{h}(k,t)\fr{j_\ell(kr)}{(kr)^2} dt.
\eeq
It is now a simple matter to calculate the tensor power spectrum using the 
orthonormality of the spin spherical harmonics and the two-point correlation 
function (\ref{hlamPS}).  The result is
\beq
C_\ell^T = \bra\alm^{T*}\alm^T\ket
    = \fr{\pi}{4}\fr{(\ell + 2)!}{(\ell - 2)!}\int\fr{dk}{k}\p_h(k)f_\ell^2(k).
\label{ClT}
\eeq
This describes the expected roughly flat plateau up to $\ell \sim 100$, 
followed by an oscillating decay.  Note the structural similarity between 
the scalar expressions (\ref{almS}) 
and (\ref{ClS}) and the corresponding tensor expressions (\ref{almT}) and 
(\ref{ClT}), with spherical harmonics in the former replaced with spin-$\lam$ 
spherical harmonics in the latter.

   Finally, we are in a position to calculate the (diagonal) tensor-scalar 
temperature anisotropy correlation.  Combining \eqs(\ref{almS}) and 
(\ref{almT}) gives
\bea
\bra\alm^{T*}\alm^S\ket
    &=& \fr{-\pi}{5\sqrt{2}}\ld[\fr{(\ell + 2)!}{(\ell - 2)!}\rd]^{1/2}
        \int\fr{dk}{k}\p_{h\R}(k)f_\ell(k)j_\ell(k\rls)\nn\\
&\times& \int d\Om_kD^{ij}e_{ij}^\lam(\kh)\sYlm{\lam}{\ell}{m}(\kh)\Ylm^*(\kh).
\label{TScorngen}
\eea
Using the explicit form \eq(\ref{zze}) for the special case of the correlation 
$D^{ij} = \hat{z}^i\hat{z}^j$, this expression becomes
\bea
\bra\alm^{T*}\alm^S\ket
    &=& \fr{-4\pi^{3/2}}{5\sqrt{30}}
        \!\!\ld[\fr{(\ell + 2)!}{(\ell - 2)!}\rd]^{1/2}
        \!\!\!\!\int\!\!\fr{dk}{k}\p_{h\R}(k)f_\ell(k)j_\ell(k\rls)\nn\\
&\times& \int d\Om_k\sYlm{\lam}{2}{0}(\kh)\sYlm{\lam}{\ell}{m}(\kh)\Ylm^*(\kh).
\eea
To evaluate the integrals over directions $\kh$, note that observations 
attempt to measure the total power at each $\ell$ mode; i.e., they attempt 
to measure
\beq
\oo{2\ell + 1}\sum_m\bra\alm^{T*}\alm^S\ket
\propto \sum_m\sYlm{\lam}{\ell}{m}(\kh)\Ylm^*(\kh).
\label{summ}
\eeq
Now, relating the spherical harmonics to the rotation matrices via
\beq
\sYlm{s}{\ell}{m}(\theta,\phi) = (-1)^m\ld(\fr{2\ell + 1}{4\pi}\rd)^{1/2}
                                D_{ms}^\ell(\phi,\theta,0)
\eeq
(I use the sign conventions of~\cite{edmonds74}), \eq(\ref{summ}) becomes
\bea
\sum_m\bra\alm^{T*}\alm^S\ket
&\propto& \sum_m D_{\lam m}^{\ell*}(0,-\theta_k,-\phi_k)
                 D_{m0}^{\ell*}(\phi_k,\theta_k,0)\nn\\
   &=& \del_{\lam0} = 0.
\label{summ0}
\eea
This result used the addition theorem for rotation matrices (see, e.g., 
\cite{edmonds74}).  Note in particular that this final result indicates that 
the expression (\ref{TScorngen}) vanishes when summed over $m$ 
{\em regardless} of the form of the coupling $D^{ij}$.  In other words, it 
is impossible to obtain a temperature anisotropy tensor-scalar 
anticorrelation to suppress the effect of tensors with the hope of 
reconciling the \Planck\ and BICEP2 limits or measurements of the 
tensor-to-scalar ratio.

\section{Conclusions}

   Note that only the {\em total} temperature power remains unsuppressed 
according to the result of \eq(\ref{summ0}).  Individual modes $\alm$ 
can be expected to be affected by the tensor-scalar correlations.  Thus we 
generically expect the appearance of statistical anisotropy, exhibited 
as off-diagonal 
correlations $\bra\alm^{T*}a_{\ell'm'}^S\ket$.  In particular, we expect 
{\em quadrupolar} anisotropy to be induced, i.e.\ couplings between $\ell$ 
and $\ell \pm 2$, due to the spin-2 character of the tensor modes.  
However, there are very tight constraints on the presence of quadrupolar 
asymmetry in the temperature anisotropies.  In particular, the \Planck\ 
measurements are consistent with zero quadrupolar asymmetry even on 
scales $\ell < 100$~\cite{planck13isostat}, where the effects of tensor 
correlations would be important.  Indeed, as mentioned in the Introduction, 
the most notable asymmetry is of dipolar character, which should not 
arise from a tensor correlation.

   As a logical, if increasingly baroque, possibility, it is worth 
mentioning that the tensor-scalar correlation tensor $D^{ij}$, which 
in this work has been assumed to be a constant, could be allowed to 
vary spatially.  With, e.g., a linear gradient in $D^{ij}$, we might 
expect that dipolar-type anisotropies could be achieved.  (Note that it 
appears to be difficult to reconcile \Planck\ with BICEP by postulating a 
gradient in $r$ across our observable volume~\cite{cdjky14}.)

   In order to exhibit analytical expressions, I used the approximations 
of the SW effect for scalars and the line of sight contribution for 
tensors.  Although these are good approximations on large scales, 
the structure of \eq(\ref{TScorngen}) should be general in that an 
improved treatment of the generation of anisotropies will change the 
detailed form of the transfer functions ($j_\ell$ for scalars and 
$f_\ell$ for tensors in my approximation), but will leave the integral 
$\int d\Om_kD^{ij}e_{ij}^\lam(\kh)\sYlm{\lam}{\ell}{m}(\kh)\Ylm^*(\kh)$ 
unchanged.  Thus the final conclusion that a suppression is not possible 
in the total temperature power should persist.

   Finally, follow-up observations by the BICEP team, and forthcoming 
polarization measurements from the \Planck\ satellite, will be crucial 
in determining whether extensions to \LCDM\ are indeed needed.  New 
measurements indicating a lower tensor-to-scalar ratio than the BICEP2 
value may reconcile temperature and polarization measurements while 
maintaining the exciting consequences of new physics.  In such a 
scenario the motivation for a suppression of large-scale temperature 
power may be removed.  Only future observations will decide whether 
the ``shouts of discrepancies'' will be silenced or will lead to a new 
view of the Universe.


\begin{acknowledgments}
This research was supported by the Canadian Space Agency.
\end{acknowledgments}

\section*{Note added}

   After the appearance of the first version of this paper, a closely related 
paper appeared~\cite{efw14}.  The results in~\cite{efw14} are in agreement 
with those presented here.

\bibliography{ts}

\end{document}